# Routes to realize the axion-insulator phase in MnBi$_2$Te$_4$(Bi$_2$Te$_3$)$_n$ family: a perspective

Yufei Zhao[1] and Qihang Liu[1,2,3*]

[1]*Shenzhen Institute for Quantum Science and Engineering (SIQSE) and Department of Physics, Southern University of Science and Technology, Shenzhen 518055, China*

[2]*Guangdong Provincial Key Laboratory for Computational Science and Material Design, Southern University of Science and Technology, Shenzhen 518055, China*

[3]*Shenzhen Key Laboratory of Advanced Quantum Functional Materials and Devices, Southern University of Science and Technology, Shenzhen 518055, China*

*liuqh@sustech.edu.cn

## Abstract

Axion, first postulated as a hypothetical particle in high-energy physics, is now extended to describe a novel topological magnetoelectric effect derived from the Chern-Simons theory in condensed matter systems. The recent discovered intrinsic magnetic topological insulators MnBi$_2$Te$_4$ and its derivatives have attracted great attention because of their potential as a material platform to realize such a quantized axion field. Since the magnetic exchange gap can bring the "half-quantized" anomalous Hall effect at the surface, an axion insulator manifests as quantum anomalous Hall and zero Hall plateau effects in the thin films. However, many puzzles about this material family remain elusive yet, such as the gapless surface state and the direct experimental evidence of the axion insulator. In this Perspective, we discuss the preconditions, manifestations and signatures of the axion-insulator phase, in the context of the development of the natural magnetic topological heterostructure MnBi$_2$Te$_4$(Bi$_2$Te$_3$)$_n$ family with various intriguing quantum phenomena. Recent theoretical and experimental efforts regarding the intrinsic magnetic topological insulators are summarized here to pave the way for this phenomenally developing field.

*Introduction*

The interplay between magnetism and topology inaugurated a new horizon in exploring exotic quantum phenomena. Rich topological phases and complex spin textures (e.g., skyrmions) have been observed over the past decades, among which the magnetic topological insulator (MTI) is one of the most boosting fields [1-5]. In a MTI, the long-range magnetic order and topological nontrivial electronic structure coexist and couple to each other below the critical temperature. As a representative example, the out-of-plane magnetization of a layered MTI breaks the time-reversal symmetry and introduces a mass term to the linear dispersion of the surface states, described by the low-energy model Hamiltonian $H_{surf} = v_F(-k_y\sigma_x + k_x\sigma_y) + m\sigma_z$, where $v_F$ denotes the Fermi velocity, $\sigma_x$, $\sigma_y$ and $\sigma_z$ the Pauli matrices, and $m$ the effective exchange field. The terms in the parentheses imply the spin-momentum locking of the massless Dirac electrons of a nonmagnetic topological insulator (TI), whereas $m\sigma_z$ modulates the bulk-edge correspondence by opening a magnetic surface gap. Characterized by 1/2 surface Chern number, such an important modification to the electronic structure provides opportunities to realize the so-called axion insulator, which is viewed as the foundation of the topological insulators and the emergent phenomena such as quantum anomalous Hall (QAH) effect and Majorana zero modes [6-13].

As a long-sought pearl in particle physics, axion is proposed in the condensed matter field theory to describe the topological magnetoelectric (TME) effect, as an additional term $S_\theta = \frac{\theta}{2\pi}\frac{e^2}{h}\boldsymbol{E} \cdot \boldsymbol{B}$, where $\boldsymbol{E}$ and $\boldsymbol{B}$ are the electromagnetic fields and $\theta$ is the pseudoscalar axion angle [4,8,9,14,15]. If the time-reversal symmetry *T* or inversion symmetry *P* is preserved, the axion field $\theta$ is quantized to 0 or $\pi$ (module $2\pi$ under periodic boundary conditions), corresponding to topological trivial and nontrivial cases, respectively. A three-dimensional gapped system could be identified as an axion insulator when it satisfies: (i) $\theta = \pi$, and (ii) gapped surface state. Such a system yields an interesting and distinct feature – half-quantized surface anomalous Hall conductance $\sigma_{xy} = \pm\frac{e^2}{2h}$ [9,16,17], which could be characterized as the direct evidence of an axion-insulator phase. Note that a 3D nonmagnetic TI fulfills the precondition (i) but not (ii). As a result, the half-quantized Hall conductance is exactly compensated by its gapless surface states, and is thus not measurable. In the two-dimensional (2D) limit, two unique quantum transport features could be probed as the manifestation of an axion insulator. Specifically, when the magnetizations of the top and bottom surfaces are parallel aligned, each of the two surfaces contributes half-quantized Hall conductance, forming a $C = 1$ (+1/2+1/2) chiral edge mode. On the other hand, when the magnetizations are antiparallel, a zero Hall plateau (ZHP) is anticipated owing to the cancellation of opposite surface Hall conductance (+1/2−1/2). In principle, one can achieve both of the two transport features in one system with different coercive fields for the top and bottom surfaces by applying a vertical magnetic field. The resultant double-step hysteresis loop, as shown in Fig. 1(a), is the outcome of an axion insulator. However, it is not the other way around as recent studies revealed that the double-step transitions with ZHP coincide with some trivial insulators and thus does not serve as sufficient evidence of the axion-insulator phase [18-21]. Several alternative approaches have also been proposed to detect axion electrodynamics signals, including the quantized optical response, charge density waves, etc [8,22-27].

Meanwhile, an ideal MTI candidate, with a high magnetic transition temperature and a large magnetic gap, is urgently needed. Huge efforts are devoted to introducing magnetism into TIs in the past few years, such as by using magnetic dopants [6,18,19,28-31] and proximity effect [32-34]. A famous successful attempt is the realization of the QAH effect in Cr-doped (Bi, Sb)$_2$Te$_3$ thin flakes, in which the magnetic gap formation and Fermi level ($E_F$) fine tuning were achieved precisely and simultaneously. However, limited by the notorious inhomogeneity [35-37] or weak coupling (tiny gap) [38,39], it is quite challenging to achieve observable QAH above ~ 1K by using these approaches, which strongly obstructs the potential applications. Such dilemma is hopefully to be changed by the recently discovered intrinsic magnetic topological materials, especially the antiferromagnetic (AFM) TI MnBi$_2$Te$_4$ and its derivatives MnTe-Bi$_2$Te$_3$ ternary system [40-44]. These stoichiometric MTIs, with an effective exchange field with better homogeneity, hold the potential to overcome the disadvantages brought by extrinsic ways to introduce magnetism, and are thus expected to elevate the realization temperature of QAH by an order of magnitude.

The born of MnBi$_2$Te$_4$-related material system has significantly advanced the field of MTI,

as reviewed by the latest literature [45-48]. In this Perspective, we aim to offer a perspective of MnBi$_2$Te$_4$-family as an ideal platform of the long-sought axion-insulator phase, with special focus on the preconditions, manifestations and direct experimental signatures. First, unlike the undoubted $\theta = \pi$ feature, the existence of the surface gap in MnBi$_2$Te$_4$-family is much more controversial. We discuss the surprising deviation between experimental measurements and theoretical predictions and provide several representative explanations of the gapless surface states. Second, a bunch of high-quality quantum transport measurements, including the QAH and ZHP effect, are reviewed as the manifestation of the axion-insulator phase. Third, we extend the focus to the exploration of the definitive signatures of the axion-insulator phase, including the chiral hinge states, disorder-induced transition, etc. Finally, we provide a brief outlook of the possible directions the prosperous field may head to.

*Gapped or gapless surface states?*

At a first glance of the crystal structure (Fig. 1(b)), MnBi$_2$Te$_4$ crystalizes in a van der Waals (vdW) structure (space group $R\bar{3}m$) with a MnTe bilayer sandwiched by Bi$_2$Te$_3$, forming a unit cell with Te-Bi-Te-Mn-Te-Bi-Te septuple layer (SL) [49-52]. Below the Néel temperature $T_N$ = 25 K, the magnetic orders within each SL are found to be parallel to the out-of-plane easy axis, but antiparallel within the adjacent SLs. Such an A-type AFM configuration enables unique thickness-dependent properties for (un)compensated magnetization [51-56]. Zhang *et al.* and Li *et al.* found that similar to other well-known tetradymites, in MnBi$_2$Te$_4$ strong spin-orbit coupling (SOC) effect leads to the band inversion between Bi-$p_z$ and Te-$p_z$ orbitals near the Fermi level at the $\Gamma$ point, while the magnetism originates from the unpaired local electron moments of the manganese ions Mn$^{2+}$ [57,58]. Hence, bulk MnBi$_2$Te$_4$ is known as the first-discovered AFM TI with a band gap of about 150 meV. Based on the prediction by Mong *et al.* [4], theoretical calculations also show that the surface state protected by $S = T\tau_{1/2}$, such as at the (100) surface, exhibits a gapless Dirac cone [40,58], while at the (001) surface with broken $S$ symmetry there is a surface gap of about 60 meV [57-59]. Within an external magnetic field about 5-10 Tesla, MnBi$_2$Te$_4$ can be tuned to a ferromagnetic (FM) state. Indeed, FM MnBi$_2$Te$_4$ locates at the transition boundary between a FM axion insulator and a Weyl semimetal, associated by a slightly lattice constant extension [20,58]. Further exploration has found that, homogeneous heterostructures MnBi$_2$Te$_4$(Bi$_2$Te$_3$)$_n$ naturally exist with alternating MnBi$_2$Te$_4$ and Bi$_2$Te$_3$ layers while maintaining topological properties [49]. Such superlattice-like MTIs exhibit termination-dependent surface states and tunable interlayer exchange coupling, such as MnBi$_4$Te$_7$, MnBi$_6$Te$_{10}$ and so on [60-68]. As *n* increases, the whole system is driven towards the Bi$_2$Te$_3$ side: the interlayer AFM coupling is gradually weakened as the separation of magnetic layers, while the global SOC strength gradually increases as the increase of Bi content. Thus, these exotic physical properties make the MnBi$_2$Te$_4$ family a promising platform for designing QAH and ZHP effects in the 2D limit [53,69].

However, in contrast to the predicted surface gap, several high-quality angle-resolved photoemission spectroscopy (ARPES) measurements observed nearly gapless surface Dirac cone in MnBi$_2$Te$_4$ bulk crystals [56,59,67,70-72] (Fig. 2(a)), despite earlier spectroscopic experiments reporting inconsistent results [40,51,54,73]. Furthermore, such a nearly vanishing gap is verified to be insensitive to temperature evolution, and is also observed at the SL termination of MnBi$_4$Te$_7$ and MnBi$_6$Te$_{10}$ [61,67,68,70]. The paradox of the surface Dirac cone and bulk magnetism sparked immediate attention. Various explanations are proposed, which can be categorized into two types, i.e., magnetic and geometric reconfigurations. Based on symmetry analysis, three types of surface spin reorientation which may give rise to linear dispersion inside the bulk gap are considered, including paramagnetism, in-plane *A*-type AFM, and *G*-type AFM [59] (Fig. 2(b)). Under these assumed magnetic structures, the $m\sigma_z$ term is dismissed or strongly suppressed. Later, other experiments yielded different microscopic evidence and further brought some new perspectives as well as questions. By using time-resolved ARPES, Nevola *et al.* observed Rashba splitting evolution inside the conduction bands with increasing temperature, which seemed to indirectly demonstrate the presence of surface ferromagnetism, yet contradicted to the persistent gapless surface Dirac cone [74]. Likewise, by using magnetic force microscopy results, Sass *et al.* revealed the robust surface AFM order persisting to the top few SLs, and suggested a possible scenario of weak coupling between topology and magnetism [71,75]. This hand waving conjecture was previously reported in MnSe/Bi$_2$Se$_3$ heterostructures [38,39], but lacks strong theoretical or experimental evidence in MnBi$_2$Te$_4$. Moreover, it seems inconsistent with the measured QAH effect in the few-layer limit

[44].

The second representative explanation of the gapless surface state is the geometric reconfiguration. Before reviewing the specific proposals, we would first like to mention the defect nature of tetradymites. It is well known that $Bi_2Te_3$ and $MnBi_2Te_4$ single crystals are typical *n*-type doped semiconductors, with the Fermi level located in the conduction band, where point defects play an important role of electron doping [76,77]. The vdW spacings between adjacent layers can serve as natural containers for impurities and intercalated atoms in the synthesis processes. Eremeev *et al.* firstly show that upon increasing vdW expansion in $Bi_2Te_3$, the Rashba-split band can be gradually shifted down from the bulk bands, which coincides with ARPES results and thus demonstrates the existence of this structural effect [78]. More importantly, in $MnBi_2Te_4$, the topological surface state is initially localized mainly in the surface SL, dominated by the magnetization from the single Mn layer. If the vdW expansion is considered, it starts to relocate inward the bulk and experiences the opposite magnetization provided by Mn atoms of the second SL [79] (Fig. 2(c)). Thus, the net magnetic moment felt by the surface state is reduced, resulting in a vanishing gap. Sun *et al.* also calculate the spatial distribution and penetration depth of the surface states from a model Hamiltonian of the bulk $MnBi_2Te_4$, indicating that the surface states are mainly embedded in the first two surface SLs [80]. In addition, based on the scanning transmission electron microscopy (STEM) observations, Hou *et al.* provided an insightful view about surface collapse [81] (Fig. 2(c)). Since the Te vacancy and Mn-Te antisite defects inevitably exist in the exfoliated surface, STEM and density functional theory (DFT) results show that a phase separation of MnTe and $Bi_2Te_3$ quintuple layer tends to occur at the surface of $MnBi_2Te_4$, probably leading to the absence of magnetic moments at some local positions.

Skipping specific speculations of reconfiguration (magnetic or geometric) that would lead to gapless surface states, a more fundamental question could be considered: whether the tendency of the gapless TSS is representative for a group of MTIs with certain features, or material dependent to $MnBi_2Te_4$. Chen *et al.* noticed a previously overlooked fact that these as-grown MTI samples are usually self-doped, where Koopmans' theorem may be applicable so as to elucidate the tendency of the surface gap [82]. If the energy gains of the doped electrons induced by closing the gap overcomes the cost of the relaxation energy of reconfiguration, a nearly gapless surface state would occur (Fig. 2(d)). The generalized theory is confirmed by DFT calculations in $MnBi_2Te_4$, $MnBi_4Te_7$ $MnBi_8Te_{13}$, yielding nice agreements with the experimental observations. Such mechanism is also expected to exist in other self-doped MTIs with small magnetic anisotropy energy, indicating the delicate interplay between magnetism and topology.

*Quantum anomalous Hall effect and zero Hall plateau*

Interestingly, despite the controversy on the magnetic surface gap, several groups has already measured QAH effect in the thin films exfoliated from $MnBi_2Te_4$ single crystals [42-44,83]. As a fundamental transport response of MTI, the QAH effect is undoubtedly inspiring news for the community. Deng *et al.* first observed a zero-field quantized Hall resistance of $\rho_{xy}$ = 0.97 $h/e^2$ in 5-SLs flakes at 1.4 K, and the quantization temperature can be elevated to 6.5 K under an external field of 7.6 Tesla [44]. Later, Liu *et al.* displayed similar field-dependent results in even-number SL flakes. As shown in Fig. 3(a), the Hall resistance $\rho_{xy}$ was initially found to be zero over a wild range around $B = 0$ T. The AFM ground state is driven to a FM state by applying a magnetic field of 6-9 T, with $\rho_{xy}$ reaching a plateau of 0.984 $h/e^2$ at 1.9 K [43]. Moreover, Ge *et al.* discovered a Chern insulator state ($C = 1$) in 7-SLs and 8-SLs devices, and an intriguing $C = 2$ state in 10-SLs devices, consistent with the predicted scenario of the FM Weyl semimetal phase [42]. Recently, Ying *et al.* also reproduced a Chern insulator state in a 5-SLs device and confirmed the chirality of edge states by using nonlocal transport measurements [83]. Except for $MnBi_2Te_4$, the QAH conductance was also obtained in $MnBi_2Te_4(Bi_2Te_3)_n$ heterostructure at 1.9 K [84]. We also noticed two outstanding work unraveled the magnetic hysteresis loop and the spin-flop transition in odd and even SLs devices by employing reflective magnetic circular dichroism (RMCD) [85,86]. For the odd layer device (5-SLs), it is found that the Hall resistance $\rho_{xy}$ and RMCD hysteresis loops are unsynchronized, indicating that the AFM state is likely a trivial magnetic insulator caused by the sample fabrication process [86]. In addition, note that almost all the probed quantized resistances have to be carried out in odd-number SLs or under a magnetic field of up to several Tesla [42-44,83,86], which poses a daunting challenge for practical applications.

Another manifestation of axion insulator is zero Hall conductivity ($\sigma_{xy}$) plateau with vanishing longitudinal conductivity ($\sigma_{xx}$), where the top and bottom surfaces exhibit opposite magnetization and $\pm \frac{e^2}{2h}$ Hall conductance [1,8]. This scenario can be traced back to a tricolor structure proposed by Wang *et al.* [87,88], whose design relies on the fact that the composing materials are well lattice matched and chemically compatible so as to ensure the magnetic exchange across the interfaces. Later, it was successfully grown by molecular beam epitaxy as shown in Fig. 3(b), where a nonmagnetic layer of (Bi, Sb)$_2$Te$_3$ is sandwiched by Cr-doped and V-doped (Bi, Sb)$_2$Te$_3$ layers [29,33,34]. The different coercive fields ($B_c$) of the two magnetic layers enable an antiparallel magnetization alignment within a certain range of external magnetic field. At 60 mK, double-step quantized transitions of $\sigma_{xy}$ and the double peaks of $\sigma_{xx}$ simultaneously occurred at $B \approx 0.2$ T and 0.8 T [34]. The expected zero Hall conductivity plateau existed between the double transitions and persisted to nearly 1 K. In MnBi$_2$Te$_4$, Liu *et al.* also observed ZHP in a 6-SL flake, which is more stable against temperature and magnetic field compared with that of the tricolor structure [43] (Fig. 3(a)). Owing to its natural symmetric structure and interlayer AFM coupling, the magnetic-field dependence curve of $\rho_{xy}$ was independent of field sweeping direction, exhibiting three distinct regimes ($C = 0, \pm 1$) but not forming a loop. As the Chern insulator state existed with the FM configuration, observed ZHP in these systems was believed to arise from the cancellation of opposite surface Hall currents (+1/2−1/2), forming an indirect evidence of axion insulator [88].

However, almost identical transport behavior was also reported at the uniformly Cr-doped (Bi, Sb)$_2$Te$_3$ samples, which indeed belong to a topological trivial case [18,19,31]. As shown in Fig. 3(c), for the six quintuple layers (QLs), the top and bottom surfaces couple with each other, resulting in a dominated hybridization gap and thus a trivial insulating phase. When applying an external field of 0.4 T, the exchange gap will overwhelm the hybridization gap, giving rise to the QAH states. Therefore, such double-step hysteresis loop is not necessarily exclusive for an axion insulator. In a thicker device (8 and 10 QLs), the ZHP between $C = \pm 1$ states disappears due to the negligible finite-size effect[19]. Another scenario of ZHP without axion-insulator phase could be achieved by the recently synthesized MnSb$_2$Te$_4$. As an isostructural compound of MnBi$_2$Te$_4$, MnSb$_2$Te$_4$ is found to exhibit both interlayer FM and AFM configurations in different growth conditions [55,89-92]. Due to a weaker SOC strength of antimony, theoretical calculations show that MnSb$_2$Te$_4$ is likely an AFM trivial insulator and a FM Weyl semimetal in its ideal crystal structure [90,93], which may also enable alternating $C = 0$ and $C = \pm 1$ states and the double-step hysteresis loop in few-layer flakes. Furthermore, plenty of ARPES experiments have observed robust gapped surface states on nonmagnetic impurities doped TI systems, whose mechanism still remains elusive [94-96]. In the 2D limits of such systems, the quantum transport behavior may also behave like that of an axion insulator. Therefore, exploring the direct experimental signature to distinguish an axion insulator and a trivial insulator is highly desirable.

### *Looking for the axion insulator phase*

The direct signature of axion-insulator phase beyond ZHP is to utilize the unique surface anomalous Hall conductivity that quantized to $e^2/2h$, and is currently confined to theoretical proposals. Chu *et al.* firstly considered a lattice model Hamiltonian of 3D TIs, in order to describe the anomalous transport of the surface electrons in the presence of magnetization [16]. They found chiral surface edge states existed at the interface of the massive top surface and the massless side surface. The Landauer-Büttiker calculations further confirmed that each of these surface edge states (hinge states) carries $e^2/2h$ quantum conductance. In such context, a comprehensive study for different topological phases based on specific material candidates would provide useful guidance for the actual experimental realization of the half-quantized surface anomalous Hall effect. Combined model Hamiltonian and DFT calculations, Gu *et al.* studied the distinctive surface anomalous Hall conductivity (AHC) features of various MnBi$_2$Te$_4$(Bi$_2$Te$_3$)$_n$ compounds in a phase diagram, including axion insulators, Weyl semimetals, 3D Chern insulators and 3D quantum spin Hall insulators [20]. They found that the local Chern marker $C_Z$ (*l*) (layer-projected Chern number) in the axion-insulator phase was well localized at the QL-terminated surface, resulting in a half-quantized surface AHC either saturated at or oscillating around $e^2/2h$, depending on the magnetic homogeneity. (Fig. 4(d)). Furthermore, they calculated the real-space local density of states of the surface and hinge states in a semi-infinite axion insulator sample [20,63]. As shown in Figs. 4(b) and 4(c), a remarkable feature of the hinge state is the asymmetric spectral weight between the left and right-moving modes, indicating its chiral nature. Therefore, the half-quantized AHC on the top and bottom surfaces will give rise to unusual chiral hinge modes embedded in the side surface Bloch states, denoted as "in-band hinge" states. These states exist within a larger energy range, and are thus expected to be an ideal spectral signature to verify the existence of the surface AHC and the axion-insulator phase.

Another proposed signature of an axion-insulator state is the recently reported 2D disorder-induced metal-insulator transition [97]. Li *et al.* introduced the random magnetic disorder into a 3D TI with top and bottom surfaces gapped by opposite magnetization, and performed a finite-size scaling analysis about the localization length, which generally increases with sample width $L$ in a metallic phase and decreases with $L$ in an insulating phase. For the weak disordered systems, as the Fermi level $E_F$ increases, the axion insulator undergoes a 2D delocalized phase transition to an Anderson insulator, and then becomes a diffusive metal. In the large disorder limit, the axion-insulator phase is gradually suppressed by the 3D critical point and eventually disappears, ending up with a 3D metal. Moreover, they showed a half-quantized surface Hall conductance plateau for the axion insulator existing at the lower energy regime, while gradually approaching zero for the Anderson insulator at the high energy regime (Fig. 4(e)). Such a universal phase transition behavior is summarized in the phase diagram shown in Fig. 4(f) and is expected to be detected in AFM axion insulator MnBi$_2$Te$_4$.

Up to now, we have focused our discussion on the "static" axion field ($\theta$ angle is independent of time). It has been proposed that the existence of dynamical axion field $\delta\theta$ (***r***, *t*) in MTIs, when magnetic fluctuations are considered, can also induce exotic responses, including axionic polariton, dynamical chiral magnetic effect, etc [15,98-101]. However, a suitable platform is lacking for a long time because most MTI candidates are centrosymmetric, resulting in a fixed axion field $\theta = \pi$. Another requirement to obtain a big dynamical axion field is that, the intrinsic MTIs must be close to the boundary of the topological phase transition. For these reasons, Zhang *et al.* proposed that Mn$_2$Bi$_2$Te$_5$ and Mn$_2$Bi$_6$Te$_{11}$-related compounds satisfy these requirements for achieving large dynamical axion field responses [102-104]. For instance, Mn$_2$Bi$_2$Te$_5$ can be viewed as intercalating two Mn-Te bilayers into the center of a Bi$_2$Te$_3$ QL, while the magnetic ground state is still *A*-type AFM, thus breaking *T* and *P* symmetry. A huge dynamical axion field factor ($1/g$) was found near the pristine $Mn_2Bi_2Te_5$ and $EuBi_2Te_5$ by slightly adjusting the SOC strength. They also proposed the nonlinear optical spectroscopy and transport measurement for the experimental detection of the dynamical axion field, which may pave an avenue towards designing future axionic devices.

*Outlook*

The discovery of intrinsic MTIs has greatly expanded our understanding of novel functional materials with unique magnetic and electronic structures. To invigorate future research in this field, we list several possible technical and scientific challenges in this section. Despite great efforts striking to theoretical proposals, the definitive signatures of the axion insulator are barely explored in experiment, especially the chiral hinge modes and the half-quantized surface AHC. The successful measurement of these effects relies foremost on a premise that the devices with desired flatness and thickness can be accurately fabricated. For the former, the scanning tunneling microscope (STM) or other advanced spatially-resolved electronic measurements, will play a significant role in examining the thickness-dependent intensity and chirality of topological hinge states [105]. For the latter, it is very promising to measure the hinge currents by using recently proposed nonlocal surface transport, but other interference factors also need to be taken into account, including the side surface states, disorder effects and thick electrodes [106,107]. Moreover, recall that the surface Hall current is the manifestation of the bulk TME effect at the surface, whose detection is a more attractive but challenging topic. It has been proposed that, several special configurations (e.g., Qi *et al.* proposed cylindrical 3D TI [9]) would give rise to a measurable TME response, but the current nanotechnology is still premature for realizing such a controllable device design.

Another direction to endeavor on is accelerating the search of MTI candidates with superior physical properties, where a combinatorial network of experimental synthesis, characterization, high-throughput computation and machine learning should be established as soon as possible [108,109]. Notably, the targeting properties of these 'inverse design' studies, coincide with the prominent features and drawbacks of $MnBi_2Te_4$ family: vdW bonding, magnetic transition temperature and nontrivial gap size. To be specific, vdW gap first avoids the appearance of interlayer dangling bonds during mechanical exfoliation, and the latter two bring QAH effects close to the conditions when materials are operated as tools and devices. After that, it is known that the MTI candidates composed of lighter elements can exhibit stable chemical bonding with great practical value, but meanwhile the global SOC strength will be weakened. By considering the coupling with other effects, such as the strong correlation effect caused by transition metals, or the Kagome lattice induced flat band structure, the topological properties are very promising to be enhanced in 2D MTIs (i.e. *d-d* band inversion). The extrinsic ways, like chemical substitution, pressure and electric field, can also do a favor to tunable exchange coupling and self-doping compensation [110-112]. It is expected that the future studies will be devoted to overcoming these puzzles and uncovering the mask of the long-sought axion insulator.

*Acknowledgements*

This work was supported by National Key R&D Program of China under Grant Nos. 2020YFA0308900 and 2019YFA0704900, Guangdong Innovative and Entrepreneurial Research Team Program under Grant No. 2017ZT07C062, Guangdong Provincial Key Laboratory for Computational Science and Material Design under Grant No. 2019B030301001, the Shenzhen Science and Technology Program (Grant No.KQTD20190929173815000) and Center for Computational Science and Engineering of Southern University of Science and Technology.

*References*

[1] Y. Tokura, K. Yasuda, and A. Tsukazaki, Nat. Rev. Phys. **1**, 126 (2019).
[2] X. Z. Yu, Y. Onose, N. Kanazawa, J. H. Park, J. H. Han, Y. Matsui, N. Nagaosa, and Y. Tokura, Nature **465**, 901 (2010).
[3] A. A. Burkov and L. Balents, Physical Review Letters **107**, 127205 (2011).
[4] R. S. K. Mong, A. M. Essin, and J. E. Moore, Phys. Rev. B **81**, 245209 (2010).
[5] L. Šmejkal, Y. Mokrousov, B. Yan, and A. H. MacDonald, Nature Physics **14**, 242 (2018).
[6] C. Z. Chang *et al.*, Science **340**, 167 (2013).
[7] Q. L. He *et al.*, Science **357**, 294 (2017).
[8] D. M. Nenno, C. A. C. Garcia, J. Gooth, C. Felser, and P. Narang, Nat. Rev. Phys. **2**, 682 (2020).
[9] X. L. Qi, T. L. Hughes, and S. C. Zhang, Phys. Rev. B **78**, 43, 195424 (2008).
[10] K. Yasuda, M. Mogi, R. Yoshimi, A. Tsukazaki, K. S. Takahashi, M. Kawasaki, F. Kagawa, and Y. Tokura,

Science **358**, 1311 (2017).

[11] A. R. Akhmerov, J. Nilsson, and C. W. J. Beenakker, Physical Review Letters **102**, 216404 (2009).

[12] B. Lian, X. Q. Sun, A. Vaezi, X. L. Qi, and S. C. Zhang, Proc. Natl. Acad. Sci. U. S. A. **115**, 10938 (2018).

[13] X.-L. Qi and S.-C. Zhang, Reviews of Modern Physics **83**, 1057 (2011).

[14] F. Wilczek, Physical Review Letters **58**, 1799 (1987).

[15] A. Sekine and K. Nomura, Journal of Applied Physics **129**, 141101 (2021).

[16] R.-L. Chu, J. Shi, and S.-Q. Shen, Phys. Rev. B **84**, 085312 (2011).

[17] N. Varnava and D. Vanderbilt, Phys. Rev. B **98**, 245117 (2018).

[18] X. Kou *et al.*, Nature Communications **6**, 8474 (2015).

[19] L. Pan *et al.*, Sci. Adv. **6**, 9, eaaz3595 (2020).

[20] M. Gu, J. Li, H. Sun, Y. Zhao, C. Liu, J. Liu, H. Lu, and Q. Liu, arXiv:2005.13943 (2020).

[21] K. M. Fijalkowski *et al.*, arXiv:2105.09608 (2021).

[22] J. Gooth *et al.*, Nature **575**, 315 (2019).

[23] L. Wu, M. Salehi, N. Koirala, J. Moon, S. Oh, and N. P. Armitage, Science **354**, 1124 (2016).

[24] Z. Wang and S.-C. Zhang, Phys. Rev. B **87**, 161107 (2013).

[25] B. J. Wieder, K.-S. Lin, and B. Bradlyn, Physical Review Research **2**, 042010 (2020).

[26] V. Dziom *et al.*, Nature Communications **8**, 8, 15197 (2017).

[27] K. N. Okada, Y. Takahashi, M. Mogi, R. Yoshimi, A. Tsukazaki, K. S. Takahashi, N. Ogawa, M. Kawasaki, and Y. Tokura, Nature Communications **7**, 6, 12245 (2016).

[28] M. Mogi, R. Yoshimi, A. Tsukazaki, K. Yasuda, Y. Kozuka, K. S. Takahashi, M. Kawasaki, and Y. Tokura, Appl. Phys. Lett. **107**, 5, 182401 (2015).

[29] M. Mogi, M. Kawamura, R. Yoshimi, A. Tsukazaki, Y. Kozuka, N. Shirakawa, K. S. Takahashi, M. Kawasaki, and Y. Tokura, Nat. Mater. **16**, 516 (2017).

[30] C.-Z. Chang *et al.*, Nat. Mater. **14**, 473 (2015).

[31] Y. Feng *et al.*, Physical Review Letters **115**, 126801 (2015).

[32] R. Watanabe *et al.*, Appl. Phys. Lett. **115**, 5, 102403 (2019).

[33] D. Xiao *et al.*, Physical Review Letters **120**, 056801 (2018).

[34] M. Mogi, M. Kawamura, A. Tsukazaki, R. Yoshimi, K. S. Takahashi, M. Kawasaki, and Y. Tokura, Sci. Adv. **3**, 5, eaao1669 (2017).

[35] S. Grauer, S. Schreyeck, M. Winnerlein, K. Brunner, C. Gould, and L. W. Molenkamp, Phys. Rev. B **92**, 5, 201304 (2015).

[36] E. O. Lachman *et al.*, Sci. Adv. **1**, 6, e1500740 (2015).

[37] I. Lee *et al.*, Proc. Natl. Acad. Sci. U. S. A. **112**, 1316 (2015).

[38] S. V. Eremeev, V. N. Men'shov, V. V. Tugushev, P. M. Echenique, and E. V. Chulkov, Phys. Rev. B **88**, 144430 (2013).

[39] V. N. Men'shov, V. V. Tugushev, S. V. Eremeev, P. M. Echenique, and E. V. Chulkov, Phys. Rev. B **88**, 224401 (2013).

[40] M. M. Otrokov *et al.*, Nature **576**, 416 (2019).

[41] E. D. L. Rienks *et al.*, Nature **576**, 423 (2019).

[42] J. Ge, Y. Liu, J. Li, H. Li, T. Luo, Y. Wu, Y. Xu, and J. Wang, National Science Review **7**, 1280 (2020).

[43] C. Liu *et al.*, Nat. Mater. **19**, 522 (2020).

[44] Y. Deng, Y. Yu, M. Z. Shi, Z. Guo, Z. Xu, J. Wang, X. H. Chen, and Y. Zhang, Science (New York, N.Y.) (2020).

[45] P. Wang, J. Ge, J. Li, Y. Liu, Y. Xu, and J. Wang, The Innovation **2**, 100098 (2021).

[46] F. C. Fei, S. Zhang, M. H. Zhang, S. A. Shah, F. Q. Song, X. F. Wang, and B. G. Wang, Adv. Mater., 25,

1904593.

[47] K. He, npj Quantum Materials **5**, 90 (2020).

[48] W. Ning and Z. Mao, APL Materials **8**, 090701 (2020).

[49] Z. S. Aliev *et al.*, J. Alloy. Compd. **789**, 443 (2019).

[50] Y. Gong *et al.*, Chinese Physics Letters **36**, 076801 (2019).

[51] A. Zeugner *et al.*, Chemistry of Materials **31**, 2795 (2019).

[52] J. Q. Yan *et al.*, Physical Review Materials **3**, 064202 (2019).

[53] M. M. Otrokov *et al.*, Physical Review Letters **122**, 6, 107202 (2019).

[54] S. H. Lee *et al.*, Physical Review Research **1**, 012011 (2019).

[55] J.-Q. Yan, S. Okamoto, M. A. McGuire, A. F. May, R. J. McQueeney, and B. C. Sales, Phys. Rev. B (2019).

[56] B. Chen *et al.*, Nature Communications (2019).

[57] J. H. Li, Y. Li, S. Q. Du, Z. Wang, B. L. Gu, S. C. Zhang, K. He, W. H. Duan, and Y. Xu, Sci. Adv. **5**, 7, eaaw5685 (2019).

[58] D. Q. Zhang, M. J. Shi, T. S. Zhu, D. Y. Xing, H. J. Zhang, and J. Wang, Physical Review Letters **122**, 6, 206401 (2019).

[59] Y. J. Hao *et al.*, Phys. Rev. X **9**, 041038, 041038 (2019).

[60] X. Wu *et al.*, Phys. Rev. X **10**, 031013 (2020).

[61] I. I. Klimovskikh *et al.*, npj Quantum Materials **5**, 54 (2020).

[62] J. Wu, F. Liu, C. Liu, Y. Wang, C. Li, Y. Lu, S. Matsuishi, and H. Hosono, Adv. Mater. **32**, 2001815 (2020).

[63] R. Lu *et al.*, Phys. Rev. X **11**, 011039 (2021).

[64] J. Z. Wu *et al.*, Sci. Adv. **5**, 9, eaax9989 (2019).

[65] C. Hu *et al.*, Nature Communications **11**, 97 (2020).

[66] X.-M. Ma *et al.*, Phys. Rev. B **102**, 245136 (2020).

[67] Y. Hu *et al.*, Phys. Rev. B **101**, 161113 (2020).

[68] N. H. Jo, L.-L. Wang, R.-J. Slager, J. Yan, Y. Wu, K. Lee, B. Schrunk, A. Vishwanath, and A. Kaminski, Phys. Rev. B **102**, 045130 (2020).

[69] H. Y. Sun *et al.*, Physical Review Letters **123**, 6, 096401 (2019).

[70] Y. J. Chen *et al.*, Phys. Rev. X **9**, 041040, 041040 (2019).

[71] H. Li *et al.*, Phys. Rev. X **9**, 041039 (2019).

[72] P. Swatek, Y. Wu, L.-L. Wang, K. Lee, B. Schrunk, J. Yan, and A. Kaminski, Phys. Rev. B **101**, 161109 (2020).

[73] R. C. Vidal *et al.*, Phys. Rev. B **100**, 121104 (2019).

[74] D. Nevola, H. X. Li, J. Q. Yan, R. G. Moore, H. N. Lee, H. Miao, and P. D. Johnson, Physical Review Letters **125**, 117205 (2020).

[75] P. M. Sass, J. Kim, D. Vanderbilt, J. Yan, and W. Wu, Physical Review Letters **125**, 037201 (2020).

[76] D. O. Scanlon, P. D. C. King, R. P. Singh, A. de la Torre, S. M. Walker, G. Balakrishnan, F. Baumberger, and C. R. A. Catlow, Adv. Mater. **24**, 2154 (2012).

[77] M.-H. Du, J. Yan, V. R. Cooper, and M. Eisenbach, Advanced Functional Materials **31**, 2006516 (2021).

[78] S. V. Eremeev, M. G. Vergniory, T. V. Menshchikova, A. A. Shaposhnikov, and E. V. Chulkov, New Journal of Physics **14**, 113030 (2012).

[79] A. M. Shikin *et al.*, Scientific Reports **10**, 13226 (2020).

[80] H.-P. Sun *et al.*, Phys. Rev. B **102**, 241406 (2020).

[81] F. Hou *et al.*, ACS Nano **14**, 11262 (2020).

[82] W. Chen, Y. Zhao, Q. Yao, J. Zhang, and Q. Liu, Phys. Rev. B **103**, L201102 (2021).

[83] Z. Ying, S. Zhang, B. Chen, B. Jia, F. Fei, M. Zhang, H. Zhang, X. Wang, and F. Song, arXiv e-prints, arXiv:2012.13719 (2020).

[84] H. Deng *et al.*, Nature Physics **17**, 36 (2021).
[85] S. Yang *et al.*, Phys. Rev. X **11**, 011003 (2021).
[86] D. Ovchinnikov *et al.*, Nano Letters **21**, 2544 (2021).
[87] J. Wang, B. Lian, and S.-C. Zhang, Phys. Rev. B **89**, 085106 (2014).
[88] J. Wang, B. Lian, X.-L. Qi, and S.-C. Zhang, Phys. Rev. B **92**, 081107 (2015).
[89] T. Murakami, Y. Nambu, T. Koretsune, G. Xiangyu, T. Yamamoto, C. M. Brown, and H. Kageyama, Phys. Rev. B **100**, 195103 (2019).
[90] Y. Liu *et al.*, Phys. Rev. X **11**, 021033 (2021).
[91] W. Ge, P. M. Sass, J. Yan, S. H. Lee, Z. Mao, and W. Wu, Phys. Rev. B **103**, 134403 (2021).
[92] S. Wimmer *et al.*, arXiv:2011.07052 (2020).
[93] L. Zhou, Z. Tan, D. Yan, Z. Fang, Y. Shi, and H. Weng, Phys. Rev. B **102**, 085114 (2020).
[94] J. Sánchez-Barriga *et al.*, Nature Communications **7**, 10559 (2016).
[95] X.-M. Ma *et al.*, Phys. Rev. B **103**, L121112 (2021).
[96] T. Sato, K. Segawa, K. Kosaka, S. Souma, K. Nakayama, K. Eto, T. Minami, Y. Ando, and T. Takahashi, Nature Physics **7**, 840 (2011).
[97] H. Li, H. Jiang, C.-Z. Chen, and X. C. Xie, Physical Review Letters **126**, 156601 (2021).
[98] A. Sekine and K. Nomura, Physical Review Letters **116**, 096401 (2016).
[99] J. Wang, B. Lian, and S.-C. Zhang, Phys. Rev. B **93**, 045115 (2016).
[100] R. Li, J. Wang, X.-L. Qi, and S.-C. Zhang, Nature Physics **6**, 284 (2010).
[101] H. Ooguri and M. Oshikawa, Physical Review Letters **108**, 161803 (2012).
[102] J. Zhang, D. Wang, M. Shi, T. Zhu, H. Zhang, and J. Wang, Chinese Physics Letters **37**, 077304 (2020).
[103] Y. Li, Y. Jiang, J. Zhang, Z. Liu, Z. Yang, and J. Wang, Phys. Rev. B **102**, 121107 (2020).
[104] H. Wang, D. Wang, Z. Yang, M. Shi, J. Ruan, D. Xing, J. Wang, and H. Zhang, Phys. Rev. B **101**, 081109 (2020).
[105] F. Lüpke *et al.*, arXiv e-prints, arXiv:2101.08247 (2021).
[106] R. Chen, S. Li, H.-P. Sun, Y. Zhao, H.-Z. Lu, and X. C. Xie, arXiv:2005.14074 (2020).
[107] Y. Li, C. Liu, Y. Wang, Z. Lian, H. Li, Y. Wu, J. Zhang, and Y. Wang, arXiv:2105.10390 (2021).
[108] Y. Xu, L. Elcoro, Z.-D. Song, B. J. Wieder, M. G. Vergniory, N. Regnault, Y. Chen, C. Felser, and B. A. Bernevig, Nature **586**, 702 (2020).
[109] N. C. Frey, M. K. Horton, J. M. Munro, S. M. Griffin, K. A. Persson, and V. B. Shenoy, Sci. Adv. **6**, 9, eabd1076 (2020).
[110] Z. Li, J. Li, K. He, X. Wan, W. Duan, and Y. Xu, Phys. Rev. B **102**, 081107 (2020).
[111] J. Shao *et al.*, arXiv:2010.11466 (2020).
[112] S. Du, P. Tang, J. Li, Z. Lin, Y. Xu, W. Duan, and A. Rubio, Physical Review Research **2**, 022025 (2020).

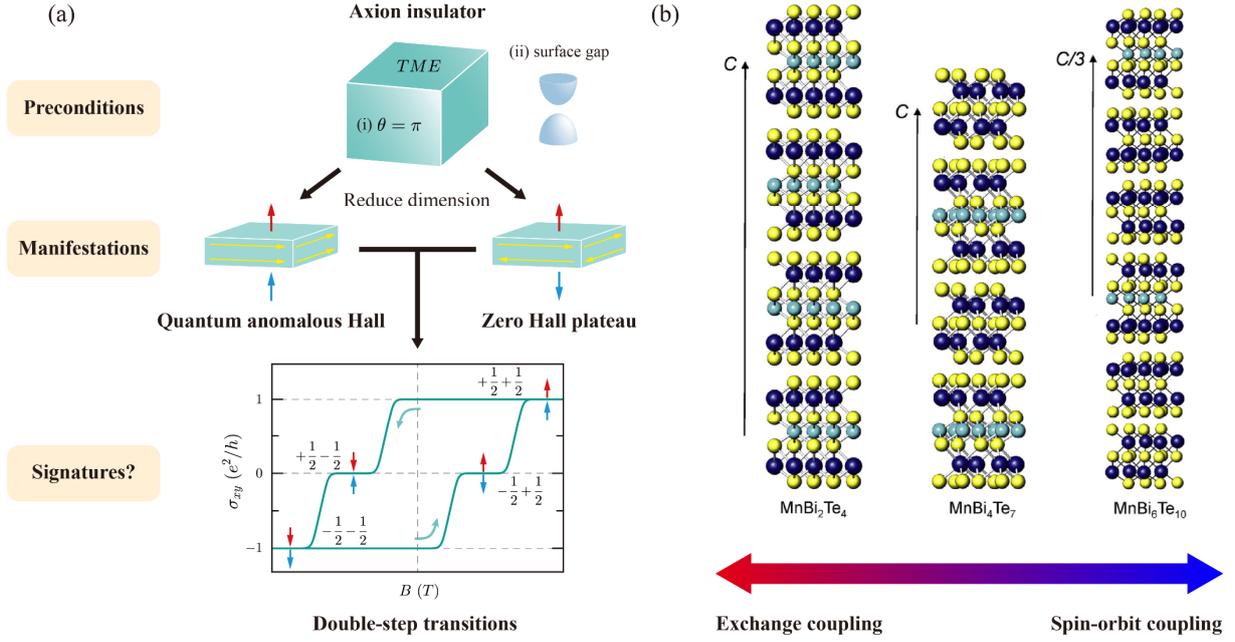

**Figure 1**: Design routes of the topological axion insulator and its material realization. (a) Preconditions: an axion insulator with the topological magnetoelectric (TME) effect need to fulfill: (i) $\theta = \pi$, and (ii) gapped surface state; Manifestations: in the two dimensional limit, the quantum anomalous Hall effect and the zero Hall plateau effect are expected to be measured when the magnetizations of the top (red arrow) and bottom (blue arrow) surfaces are aligned parallel and antiparallel, respectively. The yellow arrows denote the direction of chiral edge currents; Signatures: Magnetic field dependence of Hall conductivity $\sigma_{xy}$ with different coercive fields for the top and bottom surfaces. The cyan arrow indicates the field sweeping direction. (b) MnBi$_2$Te$_4$(Bi$_2$Te$_3$)$_n$ compounds with alternating magnetic septuple layers and nonmagnetic quintuple layers [49]. From left to right ($n$ = 0, 1, 2), the interlayer magnetic coupling decreases and the spin-orbit coupling increases, leading the system to gradually approach Bi$_2$Te$_3$ side.

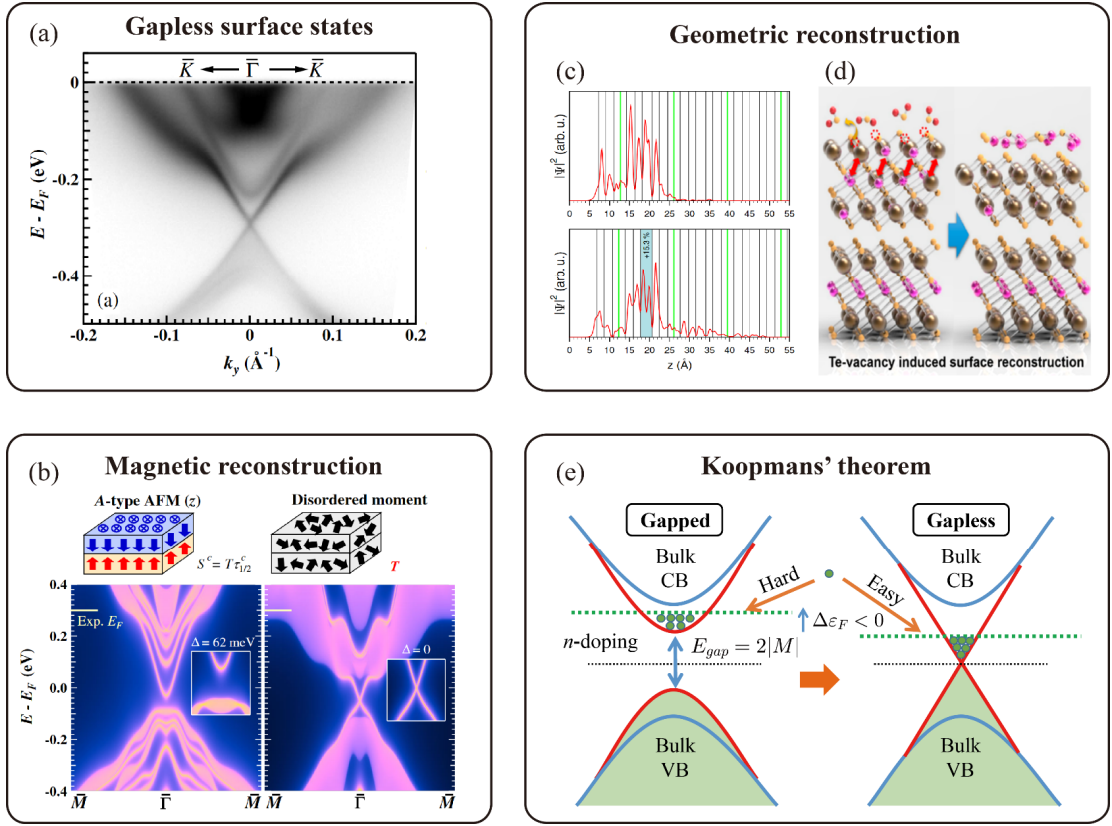

**Figure 2**: Gapless surface states of MnBi$_2$Te$_4$. (a) A typical surface Dirac cone measured by ARPES at 10 K [59]. (b) DFT calculated local density of states with A-type AFM and disordered moment configurations [59]. (c) The distribution of surface charge density with equilibrium structure (upper) and 15.3% expansion of the first van der Waals spacing (lower) [79]. (d) Schematic of the surface collapse and reconstruction in the MnBi$_2$Te$_4$ crystal induced by the Te vacancy and the Mn-Bi antisite defects [81]. (e) Schematic of Koopmans' theorem: a phase transition from gapped to gapless surface state is likely to occur in the self-doped MTI [82]. The surface (bulk) bands are represented by the red (blue) line. $\Delta\varepsilon_F$ denotes the Fermi energy difference between the gapless and gapped systems.

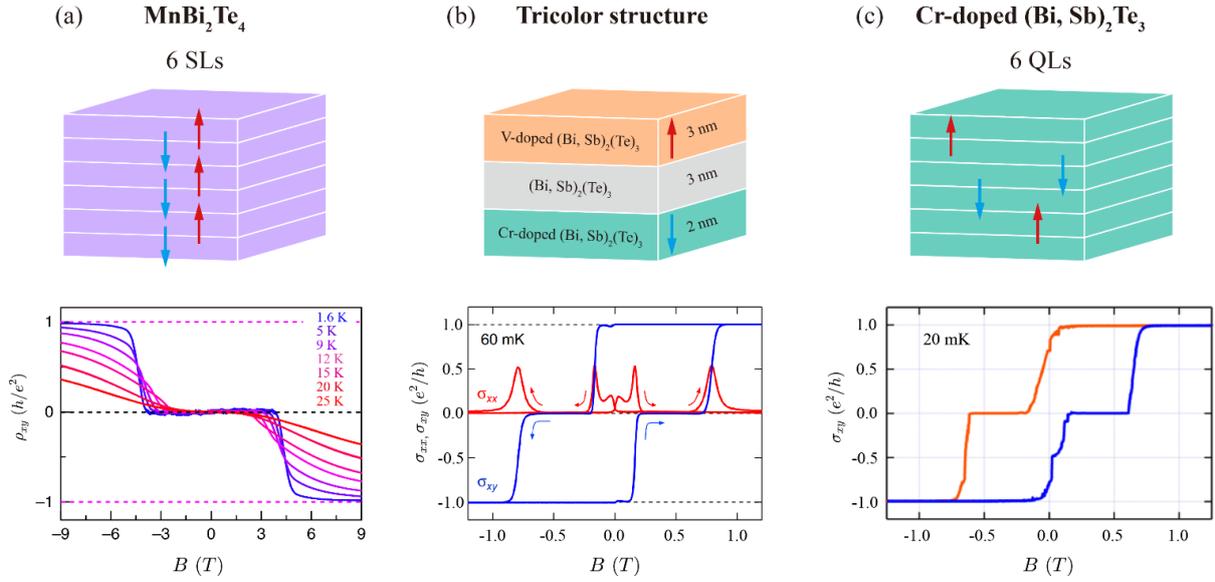

**Figure 3**: Observable double-step transitions with zero Hall plateau in (a) MnBi$_2$Te$_4$ [43], (b) tricolor structure [34], and (c) (Cr$_{0.12}$Bi$_{0.26}$Sb$_{0.62}$)$_2$Te$_3$ [19] samples. Upper panel: schematics of the thin film devices and magnetic configurations when zero Hall plateau is measured. Lower panel: Corresponding Hall resistivity/conductivity versus out-of-plane magnetic field. MnBi$_2$Te$_4$ flakes (a) naturally possess an interlayer AFM coupling at zero field. Tricolor structure enables the antiparallel magnetization in the range of 0.2 ~ 0.8 T due to the different coercive fields of V-doped and Cr-doped layers. Uniformly Cr-doped thin films (c) fail to produce a strong homogeneous magnetization under a moderate field (< 0.3 T), resulting in the exchange gap being smaller than the hybridization gap.

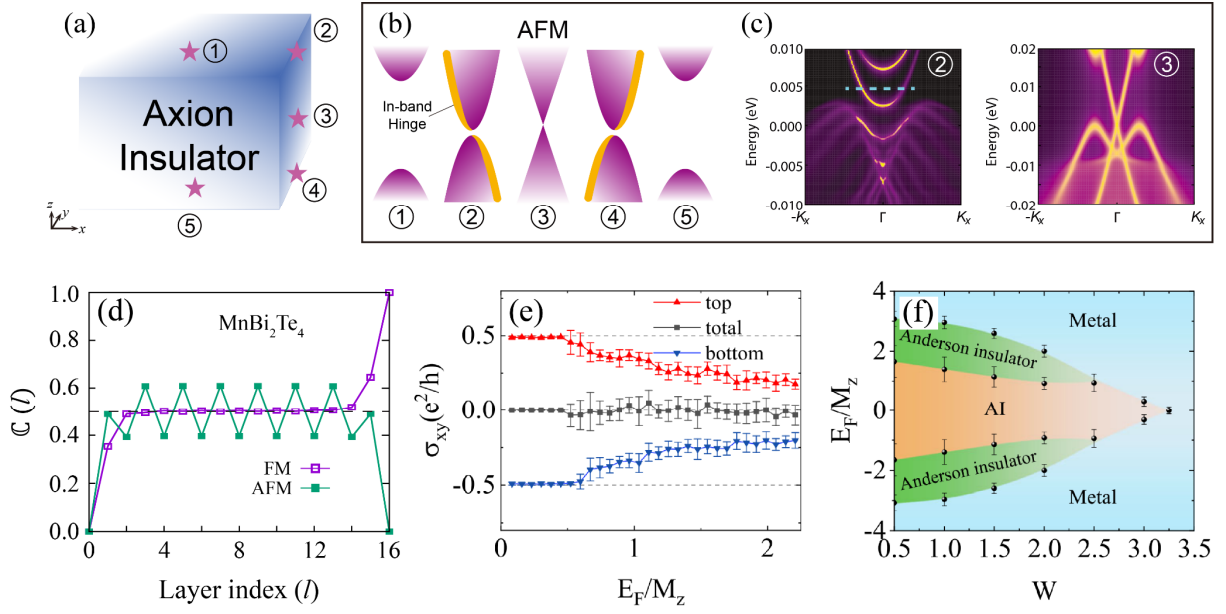

**Figure 4**: Definitive evidences of an axion insulator. (a-b) Schematic of an AFM axion insulator and the topological band spectra at spots ①-⑤, respectively [20]. (c) The hinge and side surface local density of states of AFM MnBi$_2$Te$_4$ (② and ③) [20]. (d) Integrated local Chern marker $\mathbb{C}(l)$ as a function of layer index $l$ for a 16-SL slab of MnBi$_2$Te$_4$ [20]. (e) In a weak disordered system, the Hall conductance of the top surface states, the bottom surface states and the whole sample [97]. $M_z$ denotes the surface Dirac gap. (f) Phase diagram of a disordered axion insulator in terms of relative Fermi level $E_F/M_z$ and disorder strength W [97].